\documentclass[conference]{IEEEtran}
\pdfoutput=1
\usepackage[utf8]{inputenc} % allow utf-8 input
\usepackage[T1]{fontenc}    % use 8-bit T1 fonts

\usepackage{graphicx}
\usepackage{epstopdf}
\usepackage{marginnote}
\usepackage{enumerate}
\usepackage[usenames,dvipsnames]{color}

\usepackage{amsmath}
\DeclareMathOperator*{\argmax}{argmax}
\DeclareMathOperator*{\argmin}{argmin}
\DeclareMathOperator{\sign}{sgn}
\DeclareMathOperator{\diag}{diag}
\usepackage{amssymb}
\usepackage{bm}
\usepackage{booktabs}
\usepackage{algorithm}
\usepackage{algorithmic}
\usepackage{array}
\usepackage{subfig}
\usepackage{url}
\graphicspath{{datafigures/images/}{datafigures/plots/}}

\newcommand{\V}[1]{\mathbf{#1}}
\newcommand{\bfu}{\V{u}}
\newcommand{\bfv}{\V{v}}
\newcommand{\bfw}{\V{w}}
\newcommand{\bfx}{\V{x}}
\newcommand{\bfz}{\V{z}}
\newcommand{\bfX}{\V{X}}
\newcommand{\bfU}{\V{U}}
\newcommand{\bfV}{\V{V}}
\newcommand{\bfI}{\V{I}}
\newcommand{\bfC}{\V{C}}
\newcommand{\bfB}{\V{B}}
\newcommand{\xad}{\tilde{\bfx}}%{\bfx_{\text{adv}}}

\DeclareMathOperator{\im}{im}
\newcommand{\tran}{^{\mkern-1.5mu\mathsf{T}\mkern-1.5mu}}
\newcommand{\inv}{^{\mkern-1.5mu-1}}
\newcommand{\TODO}[1]{}
\newcommand{\TODOnote}[2]{}
\newcommand{\bcomment}[1]{}

\author{{\rm Arjun Nitin Bhagoji}\\
Princeton University
\and
{\rm Daniel Cullina}\\
Princeton University
\and
{\rm Chawin Sitawarin}\\
Princeton University
\and
{\rm Prateek Mittal}\\
Princeton University
}
\begin{document}
\title{Enhancing Robustness of Machine Learning Systems via Data Transformations}
\maketitle

\begin{abstract}
We propose the use of \emph{data transformations} as a \emph{defense} against 
evasion attacks on ML classifiers. We present and investigate strategies for 
incorporating a variety of data transformations including \emph{dimensionality reduction} 
via Principal Component Analysis and data `anti-whitening' to enhance the resilience of machine learning, targeting both the classification and the training phase.
We empirically evaluate and demonstrate the feasibility of linear transformations of data as a defense mechanism against evasion attacks using multiple real-world datasets. Our key findings are that the defense is (i) effective against the best known evasion attacks from the literature, resulting in a two-fold increase in the resources required by a white-box adversary with knowledge of the defense for a successful attack, (ii) applicable across a range of ML classifiers, including Support Vector Machines and Deep Neural Networks, and (iii) generalizable to multiple application domains, including image classification and human activity classification.
\end{abstract}

\section{Introduction}\label{sec:intro}
We are living in an era of ubiquitous machine learning (ML) and artificial intelligence. 
Machine learning is being used in a number of essential applications such as image recognition~\cite{lecun2015deep}, natural language processing~\cite{collobert2011natural}, spam detection~\cite{cormack2007email}, autonomous vehicles~\cite{cirecsan2012multi,NVIDIA} and even malware detection~\cite{vsrndic2016hidost,dahl2013large}.
High classification accuracy in these settings~\cite{krizhevsky2012imagenet,taigman2014deepface,arp2014drebin} has enabled the widespread deployment of ML systems.
Given the ubiquity of ML applications, it is increasingly being deployed in \emph{adversarial} scenarios, 
where an attacker stands to gain from the failure of a ML system to classify inputs correctly. 
The question then arises: are ML systems secure in adversarial settings? 

\textit{Adversarial Machine Learning}: Starting in the early 2000s, there has been a considerable body of work~\cite{huang2011adversarial,barreno2006can,laskov2009framework} exposing the vulnerability of machine learning algorithms to strategic adversaries.
For example, \textit{poisoning attacks}~\cite{biggio2012poisoning} systematically introduce adversarial data during the \textit{training} phase with the aim of causing the misclassification of data during the test phase.
On the other hand, \textit{evasion attacks}~\cite{biggio2013evasion,papernot2016limitations,szegedy2014intriguing} aim to fool existing ML classifiers trained on benign data by adding \emph{strategic perturbations} to \emph{test inputs}. 

\textit{Evasion attacks}: In this paper we focus on evasion attacks in which the adversary aims to perturb test inputs to ML classifiers in order to cause misclassification.
Evasion attacks have been proposed for a variety of machine learning classifiers such as Support Vector Machines~\cite{biggio2013evasion,papernot2016transferability}, tree-based classifiers~\cite{papernot2016transferability,kantchelian2016evasion} 
such as random forests and boosted trees and more recently for neural networks~\cite{goodfellow2014explaining,szegedy2014intriguing,papernot2016limitations,kurakin2016adversarial,carlini2016towards,nguyen2015deep}. 
These attacks have been used to demonstrate the vulnerability of applications that use machine learning, such as 
facial recognition~\cite{mccoyd2016spoofing,sharif2016accessorize}, voice command recognition~\cite{carlini2016hidden} and \
PDF malware detection~\cite{xu2016automatically} in laboratory settings. 
Recent work also illustrates the possibility of attacks on deployed systems such as the Google video summarization API~\cite{hosseini2017deceiving}, highlighting the urgent need for defenses.
Surprisingly, it has also been shown that the evasion properties of adversarially modified data 
(for a particular classifier) persist across different ML classifiers~\cite{szegedy2014intriguing}, 
which allows an adversary with limited knowledge of the ML system to attack it. 
%Thus, it is crucial to consider the possibility of adversarial data and evasion attacks while 
%using ML systems in adversarial contexts. These vulnerabilities in ML systems make it easy for adversaries to craft inputs to meet their desired objective, which could range from fooling a self-driving car~\cite{cirecsan2012multi,NVIDIA} and making it crash to escaping fraud and malware detection~\cite{vsrndic2016hidost,dahl2013large}. 

However, very few defenses~\cite{russu2016secure,kantchelian2016evasion} exist against these attacks, and the applicability of each is limited to only certain known attacks and specific types of ML classifiers (see Section~\ref{sec:related} for a detailed description of and comparison with previous work).

\subsection{Contributions}
We propose and thoroughly investigate the use of linear transformations of data as a defense against evasion attacks. We consider powerful adversaries with \textit{knowledge of our defenses} when evaluating their effectiveness and find that they demonstrably reduce the success of evasion attacks. To the best of our knowledge, ours are the only defenses against evasion attacks with the following properties: (1) applicability across multiple ML classifiers (such as SVMs, DNNs), (2) applicability in varied application domains (image and activity classification), and (3) mitigation of multiple attack types, including strategic ones. Further, the tunability of our defense allows a system designer to pick appropriate operating points on the utility-security tradeoff curve depending on the application. 

\subsubsection{Defense} 
We propose the use of \emph{data transformations} as a defense mechanism. 
Specifically, we consider linear dimensionality reduction techniques such as Principal Component Analysis which aim to project high-dimensional data to a lower-dimensional space while preserving the most useful variance of the data~\cite{shlens2014tutorial,van2009dimensionality}.
We present and investigate a strategy for incorporating dimensionality reduction and other linear transformations of data to enhance the resilience of machine learning, targeting both the classification and training phases. 
Data transformations are applied to the training data to \textit{enhance the resilience of the trained classifier} and they significantly change the learned classifier. 
Linear data transformations are a generalization of regularization methods.
They allow us to access novel and otherwise inaccessible robustness-performance tradeoffs.

 \subsubsection{Empirical Evaluation} 
 We empirically demonstrate the feasibility and effectiveness of our defenses using: 
 \begin{itemize}
 	 \item multiple ML classifiers, such as Support Vector Machines (SVMs) and Deep Neural Networks (DNNs);
 	\item several distinct types of evasion attacks, such as an attack on Linear SVMs from Moosavi-Dezfooli et. al.~\cite{moosavi2015deepfool}, and on deep neural networks from Goodfellow et. al.~\cite{goodfellow2014explaining} and Carlini et al.~\cite{carlini2016towards}, which is the best known attack for neural networks, as well as white-box attacks targeting our defense;
 	\item a variety of real-world datasets/applications: the MNIST image dataset~\cite{lecun1998mnist} and the UCI Human Activity Recognition (HAR) dataset~\cite{anguita2013public}.
 \end{itemize} 
 
 Our key findings are that even in the face of a white-box adversary with complete knowledge of the ML system: 
\begin{itemize}
\item \textbf{Security}: the defense leads to significant increases of up to $5 \times$ in the degree of modification required for a successful attack and equivalently, reduces adversarial success rates by around $2 - 50 \times$ at fixed levels of perturbation. 
\item \textbf{Generality}: the defense can be used for different ML classifiers (and application domains) with minimal modification of the original classifiers, while still being effective at combating adversarial examples.
\item \textbf{Utility}: there is a modest utility loss of about 0.5-2\% in the classification success on benign samples in most cases. 
\end{itemize}
We also provide an analysis of the utility-security tradeoffs  as well as the computational overheads incurred by our defense. 
Our results may be reproduced using the open source code available at \url{https://github.com/anonymous}\footnote{Link anonymized for double-blind submission}.

We note that our defense does not completely solve the problem of evasion attacks, since it reduces adversarial success rates at fixed budgets, but does not make them negligible in all cases. However, since our defense is classifier and dataset-agnostic it can be used in conjunction with other defenses such as adversarial training in order to close this gap. We will explore the synergy of our defense with other techniques in future work. We hope that our work inspires further research in combating the problem of evasion attacks and securing machine learning based systems.

%The rest of the paper is organized as follows: first, in Section~\ref{sec:background} we present the necessary background on adversarial machine learning. Then, in Section~\ref{sec:defense} we describe our defense. We then set up and present our empirical evaluation in Sections~\ref{sec:exp_setup} and~\ref{sec:results} respectively. We discuss our results in Section~\ref{sec:discussion}. Finally, we provide a detailed overview of the related work in Section~\ref{sec:related} and conclude in Section~\ref{sec:conclusion}.

%%% Local Variables:
%%% mode: latex
%%% TeX-master: "NDSS"
%%% End:

\section{Adversarial machine learning} \label{sec:background}
In this section, we present the required background on 
adversarial machine learning, focusing on %(a) ML classifiers 
%such as SVMs and DNNs as well as (b) 
evasion attacks that induce misclassification by perturbing the input at test time.

\begin{figure}[t]
\centering
\subfloat[Typical test images from the MNIST dataset.
Correctly classifed as 7, 2, 1, 0 and 4 respectively.]{%
  	\includegraphics[scale=2.0]{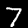}
  	\includegraphics[scale=2.0]{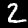}
  	\includegraphics[scale=2.0]{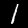}
  	\includegraphics[scale=2.0]{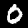}
  	\includegraphics[scale=2.0]{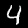}
}\hfill
\subfloat[Corresponding adversarial images obtained using the evasion attack on Linear SVMs~\cite{moosavi2015deepfool}.
Now, \textbf{misclassified as 9, 9, 3, 2 and 0} respectively.]{%
  	\includegraphics[scale=2]{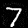}
  	\includegraphics[scale=2]{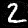}
  	\includegraphics[scale=2]{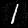}
  	\includegraphics[scale=2]{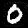}
  	\includegraphics[scale=2]{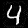}
 }
\caption{\textbf{Comparison of benign and adversarial images taken from the MNIST dataset.}}
\vspace{-10pt}
\label{fig: adv_svm_MNIST}
\end{figure}

\textbf{Motivation and Running Example}: 
We use image data from the MNIST dataset~\cite{lecun1998mnist} for our running examples.
Figure~\ref{fig: adv_svm_MNIST}(a) 
depicts example test images from the MNIST dataset that 
are correctly classified by a SVM classifier (see Section~\ref{sec:exp_setup} for details), 
while Figure~\ref{fig: adv_svm_MNIST}(b) depicts adversarially crafted 
test images (perturbed images using the evasion attack 
of Moosavi-Dezfooli et. al.~\cite{moosavi2015deepfool}), 
which are misclassified by a linear SVM. 

%\subsection{Classification using machine learning}
\textbf{Classification using machine learning:}
\label{subsec: mlclass}
We focus on \emph{supervised} machine learning, in which a classifier is trained on labeled data.
For our purposes, a classifier is a function that takes as input a data point $\bfx \in \mathbb{R}^d$ and produces an output $\hat{y} \in {C}$, where ${C}$ is the set of all categories.
The classifier succeeds if $\hat{y}$ matches the true class $y \in C$.
For example, for the MNIST dataset, $\bfx$ is a $28 \times 28$ pixel grayscale image of a handwritten digit and ${C}$ is the finite set $\{0,1,2,3,4,5,6,7,8,9\}$. 

\subsection{Attacks against machine learning systems} \label{subsec:attacks}
In this subsection we lay out the notation for the 
remainder of the paper, and describe the adversarial 
model under consideration. 
\subsubsection{Adversarial goals} \label{subsec:adv_goals} 
In each case, the adversary is given an input $\bfx$ with true class $y$ and uses an attack algorithm $A$ to produce a modified input $\xad = A(\bfx)$.
\TODO{In section~\ref{section:box}, we discuss various forms that $A$ can take.}

Attacks are relevant in the case where this input is correctly classified: $f(\bfx) = y$. The attack takes one of two forms.
\begin{itemize}
	\item \textit{Untargeted attack}: $\xad$ is misclassified ($f(\xad) \neq y$)
	\item \textit{Targeted attack}: $\xad$ is classified as a specific class $t \in C$ ($f(\xad) = t$, $t \neq y$)
\end{itemize}
Note that these attacks are equivalent for binary classifiers. Additionally, the adversarial example $\xad$ should be \emph{as similar to the benign example as possible}. Similarity is quantified using some metric on the space of examples.
As in previous work, we focus on example spaces that are vector spaces and use $p$-norms as the metric~\cite{goodfellow2014explaining,carlini2016towards}.

\subsubsection{Adversarial knowledge}
We consider three settings.
The first is the \emph{white box} setting, in which the adversary knows the classification function $f$, all its parameters and the existence of a defense, if any. 
This assumption on the adversary's capabilities is conservative as this setting corresponds to a very powerful adversary. A system with the ability to defend against attacks from an adversary with complete knowledge does not rely on security through obscurity. 

In the second setting, we consider somewhat less powerful adversaries which may be more prevalent. We call this is the \emph{classifier mismatch} setting. Here, the adversary knows the training dataset $\mathbf{X}$, the architecture of the the original classifier and the hyperparameters (i.e regularization constants etc.) used in the training of the original classifier \emph{without the defense}. Thus, the adversary is capable of training a classifier $\hat{f}$ that mimics that true classifier $f$. The adversary generates examples that are adversarial on $\hat{f}$ and submits them to $f$. Note that since the adversary is not aware of the defense measures taken, there is a mismatch between $\hat{f}$ and $f$, hence the term `classifier mismatch' for this setting.

Further, it has been shown in previous work~\cite{papernot2016transferability,szegedy2014intriguing} that adversarial examples transfer across classifiers with different architectures and hyperparameter settings, so adversaries can use their own models trained on similar datasets to construct adversarial samples for the ML system under attack. Thus, we consider a third setting, the \emph{architecture mismatch} setting\footnote{This is commonly referred to in the literature as a \emph{black-box} setting, but we use a different terminology to highlight the exact nature of the adversary's lack of knowledge.}, where the adversary is unaware of the classifier architecture being used, and just trains some $\hat{f}$ on the portion of the training data available. This is a plausible practical setting, since knowledge about the architecture and the hyperparameters of the network under attack may be difficult for the adversary to obtain.
\TODO{In this setting, the adversary may be aware of the existence of the defense...}

%Further, since even adversaries with limited access and some knowledge of dataset can \emph{infer} the classifier well enough to mount successful evasion attacks~\cite{papernot2016practical,nelson2010near,biggio2014security,papernot2016transferability},

\subsection{Evasion attacks on specific classifiers}
\begin{table*}[t]
	\centering
	\caption{\textbf{Summary of attacks on Linear SVMs and neural networks.}}
	\begin{tabular}{ccccc} \toprule
		Attack & Classifier & Constraint  & Intuition \\ \midrule
		Optimal attack on Linear SVMs ~\cite{moosavi2015deepfool} & Linear SVMs & $\ell_2$ & Move towards classifier boundary  \\ \midrule
		Fast Gradient & Neural networks & $\ell_2$ & \begin{tabular}{@{}c@{}}First order approximation to direction of smallest perturbation \end{tabular} \\ \midrule
		Fast Gradient Sign ~\cite{goodfellow2014explaining} & Neural networks & $\ell_{\infty}$ & \begin{tabular}{@{}c@{}} Constant scaling for each pixel models perception better \end{tabular}  \\ \midrule
		Carlini's $\ell_2$ attack ~\cite{carlini2016towards} & Neural networks & $\ell_{2}$ & \begin{tabular}{@{}c@{}} Iterative optimization over relaxed minimization problem \end{tabular}  \\
		\bottomrule
	\end{tabular}
	\label{table: attacks}
\end{table*}
\label{sec:known-attacks}
We now describe existing attacks from the literature for specific ML classifiers such as linear classifiers and neural networks. These are summarized in Table~\ref{table: attacks}
\subsubsection{Optimal attacks on linear classifiers}
In the multi-class classification setting for Linear SVMs, a classifier $g_i$ is trained for each class $i \in {C}$, where
\begin{align}
 g_i: \bfx \mapsto \bfw_i ^T \bfx+b_i.
\end{align}
The final classifier $f$ assigns $\bfx$ to the class $f(\bfx) = \argmax_{i \in {C}} g_i(\bfx)$.
Given that the true class is $y \in C$, the objective of an untargeted attack is to find the closest point $\xad$ such that $f(\xad) \neq y$.

From~\cite{moosavi2015deepfool}, we know the optimal attacks on affine multi-class classifiers under the $\ell_2$ metric.
This attack finds $\xad$ that minimizes $\|\xad - \bfx \|_2$ subject to the constraint $f(\xad) \neq t$.
For $\bfx$ such that $f(\bfx) = y$ and targeted class $t \in C$, let $\bfz_{t,y} = \bfw_t - \bfw_y$.
Observe that $g_t(\bfx) - g_y(\bfx) = \bfz_{t,y}\tran\bfx$.
Then, the adversarial example 
\begin{align}
\xad_t =\bfx - \frac{ (\bfz_{t,y}\tran\bfx) \bfz_{t,y}}{\|\bfz_{t,y}\|^2_2}
\end{align}
satisfies $\bfz_{t,y}\tran\xad_t = 0$.
The minimum modification required to cause a misclassification as $t$ is 
\begin{equation}
\epsilon_t 
= \|\bfx - \xad_t\|_2 
= \frac{\| \bfz_{t,y} \bfz_{t,y}\tran\bfx \|_2}{\|\bfz_{t,y}\|_2^2} 
= \frac{\bfz_{t,y}\tran\bfx}{\|\bfz_{t,y}\|_2}.
\end{equation}
Thus the optimal choice of $\xad$ for an untargeted attack is $\xad_k$, where $k = \argmin_t \epsilon_t$.

\subsubsection{Gradient based attacks on neural networks}
The Fast Gradient Sign (FGS) attack is an efficient attack against neural networks introduced in~\cite{goodfellow2014explaining}.
This attack is for the $\ell_{\infty}$ metric.
Adversarial examples are generated by adding adversarial noise proportional to the sign of the gradient of the loss function $J(\bfx,y,\theta)$.
Here, $\bfx$ is the example, $y$ is the true class, and $\theta$ is the network weight parameters.
Concretely,
\begin{equation}
\xad(\eta)_i = \bfx_i + \eta \sign (\nabla_{\bfx} J(\bfx,y,\theta))_i.
\end{equation}
The gradient can be efficiently calculated using backpropagation.
The parameter $\eta$ controls the magnitude of the adversarial perturbation, similar to $\epsilon$ for the attack on Linear SVMs:
\begin{align}
\Vert \bfx - \xad(\eta) \Vert_{\infty}=\max_i \vert \eta \sign (\nabla_{\bfx} J(\bfx,y,\theta))_i \vert =\eta.
\end{align}
See Figure~\ref{fig:adv_fsg_images} in the Appendix for images modified with a range of $\eta$.

The FGS attack and the attack on Linear SVMs are constrained according to different norms.
To facilitate a comparison of the robustness of classifiers as well as the effectiveness of our defense across them, we propose a modification of the FGS attack which is constrained by the $\ell_2$ norm.
Denoting this as the Fast Gradient (FG) attack, we define the adversarial examples to be equal to
\begin{equation}
\xad(\epsilon)= \bfx+ \epsilon  \frac{\nabla_{\bfx} J(\bfx,y,\theta)}{\| \nabla_{\bfx} J(\bfx,y,\theta) \|_2}.
\end{equation}
For the FG attack, $\epsilon$ is the $\ell_2$ norm of the perturbation. 

\subsubsection{Optimization-based attacks on neural networks}
The direct optimization based formulation of adversarial sample generation for a classifier $f$ is
\begin{align}
\min \quad &d(\xad, \bfx),\\
\text{s.t.} \quad &f(\xad) \neq f(\bfx) \nonumber \\
 &\xad \in C \nonumber,
\end{align}
where $d$ is an appropriately chosen distance metric and $C$ is the constraint on the input space. Since the constraint $f(\xad) \neq f(\bfx)$ in the above optimization problem is combinatorial, various related forms of this minimization problem have been proposed~\cite{szegedy2014intriguing,carlini2016towards,moosavi2015deepfool}. We focus on the relaxation studied by Carlini and Wagner~\cite{carlini2016towards}:
\begin{align}
\min \quad & d(\xad, \bfx) + \lambda \ell(\xad, f),\\
\text{s.t.} \quad & \xad \in C \nonumber.
\end{align}
Carlini and Wagner investigate a variety of loss functions $\ell(\cdot)$ as well as methods to ensure the generated adversarial sample stays within the input space constraints. In our experiments with neural networks, for untargeted attacks we the use a loss function, $\max(Z(\xad)_o-\max\{Z(\xad): i \neq o\}, -\kappa)$, where $o$ is the original class of the input, $Z(\cdot)$ represents the output of the neural network before the softmax layer and $\kappa$ represents the confidence parameter. The distance metric used is the $\ell_2$ norm since it is found to perform the best.

We evaluate the effectiveness of our defense against this state-of-the-art attack in Section~\ref{sec:results}.

%%% Local Variables:
%%% mode: latex
%%% TeX-master: "NDSS"
%%% End:

\section{Data transformations as a defense}\label{sec:defense}
In the previous section, we have seen that ML classifiers are vulnerable to a variety of different evasion attacks.
Thus, there is a clear need for a defense mechanism that is effective against a variety of attacks, since a priori, the owner of the system has no knowledge of the range of possible attacks.
Further, finding a defense that works across multiple classifiers can direct us to a better understanding of why ML systems are vulnerable in the first place.

In this section, we present a defense that is resilient to attacks from the literature in the mismatch setting, and remains effective even in the presence of a white-box adversary with knowledge of the defense.
Further, the defense makes multiple types of ML classifiers operating in different application scenarios more robust as shown by our results in Section~\ref{sec:results}.
The defense is based on linear transformations of data, including linear dimensionality reduction.

\subsection{Overview of defense}
\label{section:defense-overview}
The dimension of the data is $d$ and the training data is a $d \times n$ matrix $\V{X}$, so each example is a column.
We assume the data is centered, i.e. $\bfX \V{1} = \V{0}$ where $\V{1} \in \mathbb{R}^n$ is the vector of all ones and $\V{0}\in \mathbb{R}^d$ is the vector of all zeros.
The set of data classes is $C$ and the classifier in use is $f : \mathbb{R}^d \to C$.

In our defense, we leverage linear transformations of the data to make the classifier more resilient by modifying the training phase.
In the first step of our defense, an algorithm selects a linear transformation such as dimensionality reduction based on properties of the data distribution.
Then, the training data $\bfX$ is transformed and a new classifier $f$ is then trained on the transformed \textit{training set}.
In the classification phase, all inputs are transformed in the same way before being provided to the classifier.

\begin{algorithm}[H]	
  \caption{\textsf{LTtrain}}
  \label{alg:LTtrain}
  % \noindent\makebox[\textwidth]{\hrulefill{0.4pt}}
  \begin{algorithmic}[1]
    \REQUIRE $\bfX$, \textsf{Train}, \textsf{Select}
    \ENSURE $f$
    \STATE Select the linear transformation $\bfB = \textsf{Select}(\bfX)$ 
    \STATE Compute the transformed training set $\bfB \bfX$
    \STATE Train classifier $f^{\text{aux}}=\textsf{Train}(\bfB \bfX)$
    \STATE Let $f = (\bfx \mapsto f^{\text{aux}}(\bfB\bfx))$ 
  \end{algorithmic}
\end{algorithm}

The additional inputs to Algorithm~\ref{alg:LTtrain} are:
\begin{itemize}
  \item \textsf{Select}: The algorithm used to select a linear transformation of the data based on some properties of the data.
This may be a specialization of a more general algorithm to specific parameters: e.g. $\textsf{Select} = \textsf{TopPrincipalComponents}(k)$. 
  \item \textsf{Train}: This is the algorithm used to train classifiers of the desired class.
In general, this will be a specialization of more general training algorithm to specific parameters.
For example, \textsf{Train} might produce a neural network via Stochastic Gradient Descent~\cite{goodfellow2016deep} starting from the untrained classifier network $f^0$ using training parameters $\bm{\theta}$.
\end{itemize}

At first glance, this approach may seem futile, because the initial layer of many machine learning classifiers, including SVMs and neural networks, is a linear function.
However, although these classifiers are already capable of applying any linear transformation to the data that the training procedure finds to be useful, the standard training process does not optimize for adversarial robustness and does not choose to make these transformation, even though they are available.
Thus, an \textit{explicit linear transformation of the data is capable of significantly changing the learned classifier} and a carefully selected transformation can lead to beneficial changes.

\subsection{Effect on Support Vector Machines}
To motivate Algorithm~\ref{alg:LTtrain}, we examine in detail the case where \textsf{Train} produces a linear classifier by learning a support vector machine.

Learning a SVM \cite{scholkopf_learning_2002} that can classify data points from two classes, $y_i \in \{-1,1\}$, involves finding an affine function $f(\bfx) = \bfw\tran\bfx + b$ that minimizes the following loss function
\begin{equation}
  L(\bfX;\bfw,b) = \frac{1}{2}\bfw\tran\bfw + \sum_{i} \max(-1,y_i(\bfw\tran\bfx_i + b)). \label{svm-loss}
\end{equation}

If we use Algorithm~\ref{alg:LTtrain} and apply an invertible linear transformation $\bfB$ to the training data, we will learn an alternative function $g^{\text{aux}}(\bfx) =  \bfu\tran\bfx + b'$ that minimizes
\begin{equation}
  L(\V{AX};\bfu,b') = \frac{1}{2}\bfu\tran\bfu + \sum_{i} \max(-1,y_i(\bfu\tran\bfB\bfx_i + b)). \label{transformed-loss}
\end{equation}
Our actual classifier will be $g(\bfx) = g^{\text{aux}}(\bfB\bfx) = \bfu\tran \bfB \bfx + b'$.
Let $\bfw = \bfB\tran\bfu$ and rewrite \eqref{transformed-loss} as 
%For any $\bfu$, the loss $L(\V{AX};\bfu,b')$ is equal to the
\begin{multline}
  \frac{1}{2}\bfu\tran \bfB \bfB\inv (\bfB\inv)\tran \bfB\tran \bfu + \sum_{i} \max(-1,y_i(\bfu\tran \bfB \bfx_i + b)) \\
  = \frac{1}{2}\bfw\tran (\bfB \bfB\tran)\inv \bfw + \sum_{i} \max(-1,y_i(\bfw\tran \bfx_i + b)) . \label{quadratic-form-loss}
\end{multline}
Selecting the $\bfu$ that minimizes \eqref{transformed-loss} and composing it with $\bfB\tran$ is equivalent to directly selecting the value of $\bfw$ that minimizes \eqref{quadratic-form-loss}.
Thus applying an invertible linear transformation to the data is equivalent to modifying the quadratic form that appears in the regularization term of the SVM loss function.

\textbf{Regularization:}
A standard generalization of \eqref{svm-loss} multiplies the $\frac{1}{2}\bfw\tran\bfw$ by a regularization parameter $\lambda$.
This corresponds to the simplest possible linear transformation of the data: multiplication by a constant.
Explicitly, we have $\bfB = \frac{1}{\sqrt{\lambda}}\bfI$. 
Thus, ordinary regularization of SVMs fits neatly into this framework. 
However, more general linear transformations provide us with significantly more flexibility to modify the regularization constraint and allow us to \textit{access novel and otherwise inaccessible robustness-performance tradeoffs}.

\textbf{Singular linear transformations:}
What happens if $\bfB$ is not invertible?
In this case, $\bfB\tran \bfu$ is a member of $\im{\bfB\tran}$ by definition.%
\footnote{The image of the operator $\bfB\tran$ is the vector space of linear combinations of columns of the matrix $\bfB\tran$.}
%orthogonal to the null space of $\bfB$ because $\bfB\bfx = \V{0}$ implies $\bfu\tran \bfB \bfx = 0$.
Then $\bfw = \bfB\tran \bfu$ minimizes
\[
  \frac{1}{2}\bfw\tran (\bfB \bfB\tran)^{+} \bfw + \sum_{i} \max(-1,y_i(\bfw\tran \bfx_i + b)) \label{pseudo-loss}
\]
over $\bfw \in \im{\bfB\tran}$, where $(\bfB \bfB\tran)^{+}$ is the Moore-Penrose pseudoinverse of $\bfB$.
\footnote{Because $\bfB \bfB\tran$ is a symmetric matrix, it has a spectral decomposition $\bfB \bfB\tran = \bfV \Lambda \bfV\tran$ where $\Lambda$ is diagonal. Then $(\bfB \bfB\tran)^{+} = \bfV \Lambda^{+} \bfV\tran$ where $\Lambda^{+}$ is diagonal, $(\Lambda^{+})_{ii} = \Lambda_{ii}^{-1}$ if $\Lambda_{ii} \neq 0$, and $(\Lambda^{+})_{ii} = 0$ otherwise.}
\cite{penrose_best_1956}.
In addition to modifying the costs assigned to each weight vector, applying a non-invertible transformation $\bfB$ to the data rules out some choices of $\bfw$ completely.
Alternatively, one can think of the regularization term as assigning an infinite cost to each $\bfw \in \ker{\bfB}$ and thus to each $\bfw \not\in \im{\bfB\tran}$.%
\footnote{The kernel of the operator $\bfB$ is the space of vectors $\bfx$ such that $\bfB \bfx = \V{0}$.
$\ker{\bfB}$ is orthogonal to $\im{\bfB\tran}$.} 
\TODO{
Is there any advantage to this argument?
I think it proves move, not only is $\bfB\tran \bfw$ in $\im A\tran$, $\bfw$ is in $\im A$, thus the values of the two loss functions are equal.
Argument: write $\bfw = \bfu + \bfv$ where $\bfu \in \im A$ and $\bfv \in \ker A\tran$
Then $\bfv$ does not affect data dependent part of loss but does affect regularization, so it is zero.

One important example of this is Linear SVMs trained using $\ell_2$ regularization.
In this case, Algorithms~\ref{alg:DRtrain} and \ref{alg:altDRtrain} will clearly select linear functions with the same behavior on each of the training points.
This extends by linearity to the subspace of $\mathbb{R}^n$ containing all of the projected data points.

Among the linear functions that achieve this behavior, regularization will force Algorithm~\ref{alg:altDRtrain} to select the function with the smallest weight vector in the $\ell_2$ norm.
This will be the linear function that is invariant on all vectors orthogonal to the subspace containing the projected data.
In practice, we prefer Algorithm~\ref{alg:DRtrain} because it operates on lower dimensional representations of the vectors and thus is more computationally efficient.
}

\textbf{Expressivity:}
For a fixed collection of data points, it is possible to find a linear transformation that results in the selection of essentially any hard-decision classifier positively correlated with the true labels.
Observe that this is the opposite of the naive fear described in Section~\ref{section:defense-overview} that linear transformations should have no effect on the learned classifier: the choice of linear transformation might influence the final classifier structure too much! 
Because of this, it is essential to select linear transformations in a systematic matter.

\subsection{Defense using PCA}
Several of the choices of $\textsf{Select}$ that we will use in Algorithm~\ref{alg:LTtrain} are based on Principal Component Analysis.
\subsubsection{PCA in brief}
PCA~\cite{shlens2014tutorial} is a linear transformation of the data that identifies so-called `principal axes' in the space of the data, which are the directions in which the data has maximum variance, and projects the original data along these axes.
The dimensionality of the data is reduced by choosing to project it along $k$ principal axes.
The choice of $k$ depends on what percentage of the original variance is to be retained.
Intuitively, PCA identifies the directions in which the `signal', or useful information in the data is present, and discards the rest as noise. 

Concretely, let the data samples be column vectors $\bfx_i \in \mathbb{R}^d$ for $i \in \{1, \ldots, n\}$, let $\bfX$ be the $d \times n$ matrix of centered data samples.
The principal components of $\bfX$ are the normalized eigenvectors of its sample covariance matrix $\bfC = \bfX \bfX\tran$.
More precisely, because $\bfC$ is positive semidefinite, there is a decomposition $\bfC = \bfU \Lambda \bfU\tran$ where $\bfU$ is an orthogonal matrix, $\Lambda = \diag(\lambda_1,\ldots,\lambda_d)$, and $\lambda_1 \geq \ldots \geq \lambda_d \geq 0$.
In particular, $\bfU$ is the $d \times d$ matrix whose columns are unit eigenvectors of $\bfC$.
The eigenvalue $\lambda_i$ is the variance of $\bfX$ along the $i^{\text{th}}$ component.

Each column of $\bfU\tran\bfX$ is a data sample represented in the principal component basis.
Let $\bfX_k$ be the projection of the sample data in the $k$-dimensional subspace spanned by the $k$ largest principal components.
Thus $\bfX_k=\bfU \bfI_k \bfI_k\tran \bfU\tran \bfX$, where $\bfI_k$ is a $d \times k$ rectangular diagonal matrix.
%with ones in the first $k$ positions and zeros in the last $d-k$.
The amount of variance retained is $\sum_{i=1}^k \lambda_i$, which is the sum of the $k$ largest eigenvalues.

\subsubsection{Implementing the defense}
There are two choices of a linear transformation that keep the $k$ largest principal components: $\bfB = \bfI_k\tran \bfU\tran$, which is a $k \times d$ matrix, and $\bfB = \bfU\tran \bfI_k \bfU$, which is a $d \times d$ matrix.
For some choices of \textsf{Train} including SVMs trained using \eqref{svm-loss}, these choices of $A$ are equivalent, i.e. they will output identical classifiers given the same inputs.
The choice $\bfB = \bfI_k\tran \bfU\tran$ allows for more efficient training because representation of the data is more compact.
The choice $\bfB = \bfU \bfI_k \bfI_k\tran \bfU\tran$ makes is it easier to compare the reduced dimension data to the original data.

The complexity analysis for the PCA-based defense is in Section~\ref{sec:complexity} (c.f. Appendix).

\subsection{Intuition behind the PCA defense}
\label{section:intuition}

We will give some intuition about why dimensionality reduction should improve resilience for SVMs.
We discuss the two-class case for simplicity, but the ideas generalize to the multiple class case.
The core of a linear classifier is a function $g(\bfx) = \bfw\tran\bfx + b$.
Both $\bfx$ and $\bfw$ can be expressed in the principal component basis as $\bfU\tran\bfx$ and $\bfU\tran\bfw$.
We expect that many of the principal components with the largest coefficients in the weight vector, $|(\bfU\tran\bfw)_i|$, to correspond to small eigenvalues $\lambda_i$.

\begin{figure}
  \includegraphics[scale=0.3]{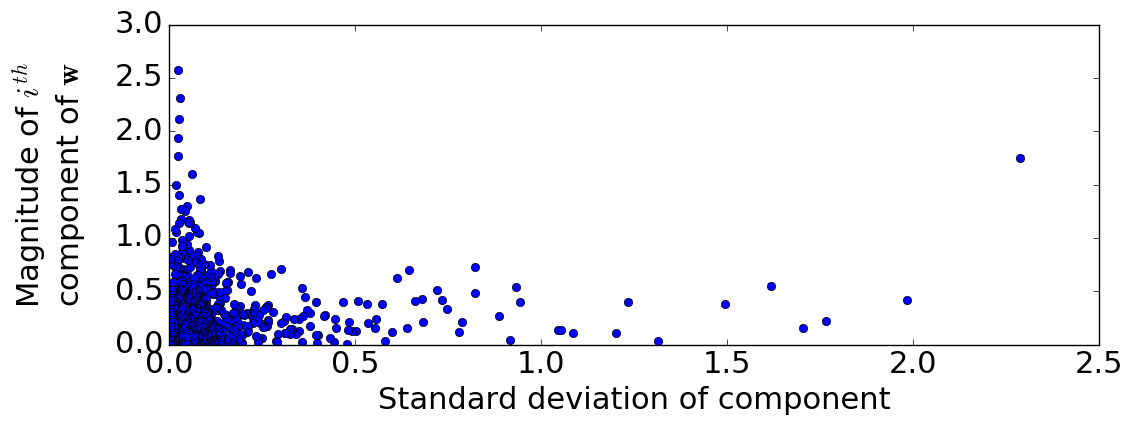}

  \caption{
    \textbf{Magnitudes of the coefficients of the weight vector $\bfw$ of a linear SVM in the principal component basis}.
    On the horizontal axis, we have $\sqrt{\lambda_i}$.
    On the vertical axis, $|(\bfU\tran\bfw)_i|$.
    The classifier is trained on the original MNIST data.}

  \label{fig:w-scatter}
\end{figure}

The reason for this is very simple: in order for different principal components to achieve the same level of influence on the classifier output, $|(\bfU\tran\bfw)_i|$ must be proportional to $1/\sqrt{\lambda_i}$.
To take advantage of the information in a component with small variation, the classifier must use a large coefficient.
Of course, the principal components vary in their usefulness for classification.
However, among the components that are useful, we expect a general trend of decreasing coefficients of $(\bfU\tran\bfw)_i$ as $\sqrt{\lambda_i}$ increases. 

Figure~\ref{fig:w-scatter} validates this prediction.
Many of the principal components with very low variances have large coefficients in $\bfw$.
As variance increases, the coefficients tend to decrease.
The exception to the trend is the first principal component, but this is not surprising.
The first principal component\footnote{on the top right in each plot} is by far the most useful source of classification information because it is strongly aligned with the difference of the class means.
Consequently it does not fit the overall trend and actually has the largest coefficient.
However, among the other components, there is a mixture of cross-class and in-class variation and the trend holds.

%Furthermore, we expect higher variance components to contain better information than lower variance components.
%By better information, we mean components that generalize better from the training set to the test set, or equivalently components that mostly correspond to underlying features of the classes rather than spurious correlations driven by the specific training examples.

\textbf{Effect on robustness:}
Since the optimal attack perturbation for a linear classifier is a multiple of $\bfw$, the \emph{principal components with large coefficients are the ones that the attacker takes advantage of}.
The defense denies this opportunity to the attacker by forcing the classifier to assign no weight to the low variance components.
This significantly changes the resulting $\bfw$ that the classifier learns.
The classifier loses access to some information, but accessing that information required large weight coefficients, so the attacker is hurt far more.
\emph{Thus, by using only high variance components, the classifier gains significant adversarial robustness for the loss of a small amount of classification performance.}

\begin{figure}
  \centering
  \subfloat[$k = 331$]{
    \includegraphics[scale=0.3]{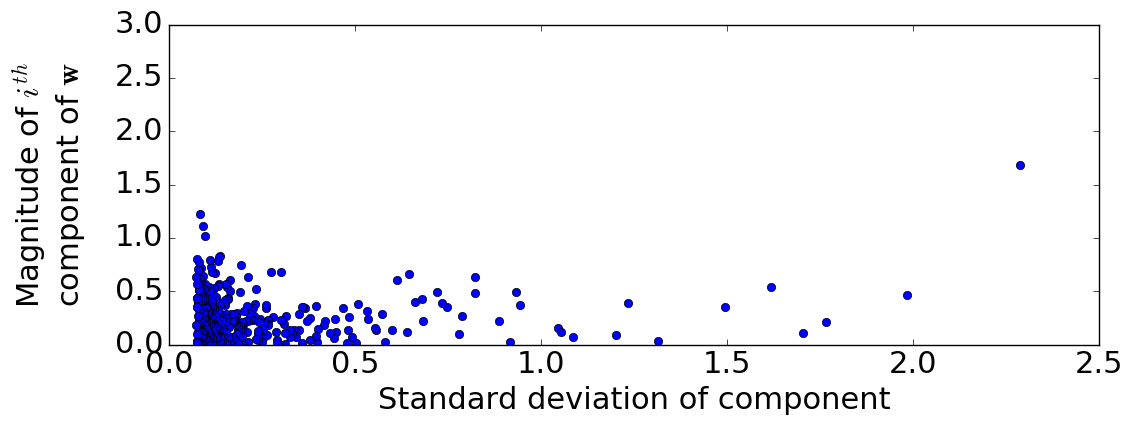}
  }

  \subfloat[$k = 100$]{
    \includegraphics[scale=0.3]{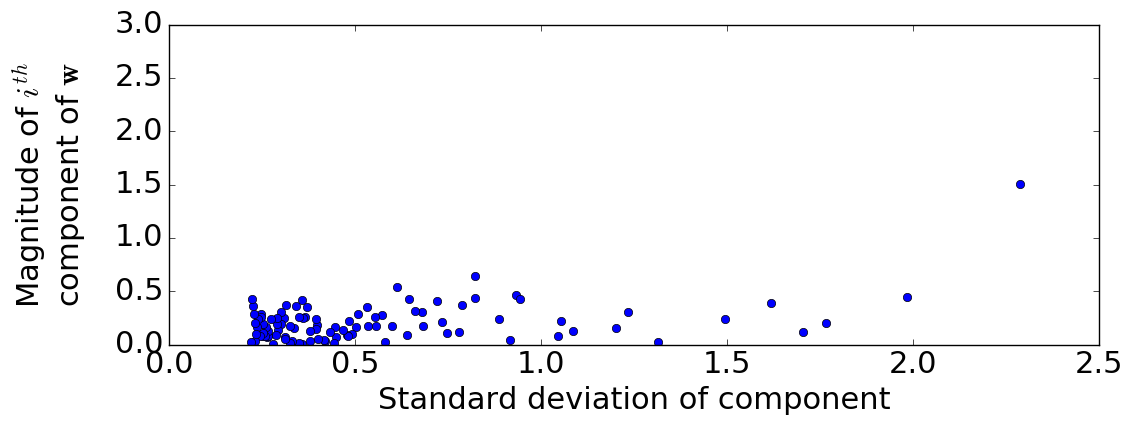}
  }

  \caption{\textbf{The magnitudes of the coefficients of the weight vector $\bfw$ of a linear SVM in the principal component basis}.
On the horizontal axis, we have $\sqrt{\lambda_i}$.
On the vertical axis, $|(\bfU\tran\bfw)_i|$.
The classifiers are trained on the MNIST data projected onto the top $k$ principal components.}

  \label{fig:w-scatter-pca}
\end{figure}

Figure~\ref{fig:w-scatter-pca} shows the coefficient magnitudes for classifiers trained on data projected onto the top $k$ principal components.
Observe that eliminating the low variance principal components mostly removes the relationship between the variance of a component and the corresponding coefficient of $\bfU\tran \bfw$.

\subsection{Other linear transformations and classifiers}\label{subsec: antiwhiten}
\subsubsection{Anti-whitening}
Now we will discuss another linear transformation based on the principal components of the training data that can confer additional robustness.
As before, we have $\bfC = \bfX \bfX\tran = \bfU \Lambda \bfU\tran$.
The linear transformation which we call `anti-whitening' is accomplished by selecting $\bfB = \Lambda^{\frac{c}{2}} \bfU$ and $\bfB = \bfU \Lambda^{\frac{c}{2}} \bfU$ for some $c > 0$.
In  our experiments, we use the former but the latter is conceptually easier to work with.
Anti-whitening exaggerates the disparity between the variances of the components.
In the full rank case, the quadratic form introduced in the SVM loss is $\bfC^{-c}$.
It serves as a softer alternative to completely eliminating low variance principal components: the hard cutoff $\bfI_k \bfI_k\tran$ is replaced with the gradual penalty $\Lambda^{\frac{c}{2}}$.
Low variance components are still available for use, but the price of accessing them is increased.

\bcomment{
The standard process of whitening accomplished by setting $c=-1$ runs counter to the intuition that we discussed in Section~\ref{section:intuition}.
It places all principal components on an equal footing, penalizing the use of low variance components less than in the initial loss function.
The aim of whitening is to find a rescaling of the principal components that results in an identity sample covariance: $\bfB \bfX \bfX\tran \bfB\tran = \bfI$.
As with PCA dimensionality reduction, there are two choices of linear transformation.
If $\bfC$ is full rank, this is accomplished by $\bfB = \Lambda^{-\frac{1}{2}} \bfU$ and $\bfB = \bfU \Lambda^{-\frac{1}{2}} \bfU$.
In experiments, we use the former, but the latter is conceptually easier to work with.
If $\bfC$ has rank $k < d$, we include only the $k$ nontrivial eigenvalues in $\Lambda$.
Then this choice of $A$ achieves the best feasible result: uniform variance within the subspace spanned by the data. 

Observe that whitening corresponds to the using the quadratic form $(\bfB \bfB\tran)\inv = (\bfU \Lambda\inv \bfU\tran)\inv = \bfU \Lambda \bfU\tran = \bfC $ in the SVM loss.
This is the same quadratic term that appears in the squared loss function used in standard linear regression.
}

\subsubsection{Neural networks}
Neural networks are more complicated due to the non-uniqueness of local minima in the associated loss function and the larger variety of regularization methods that are employed.
At first glance, it may seem that adding a linear layer as the first layer of a neural network may provide the same benefits as PCA-based dimensionality reduction.
However, the training process does not optimize for robustness, so in practice the linear layer that is learned does not have the desired effect\footnote{In fact, Gu and Rigazio~\cite{gu2014towards} found that even non-linear denoising layers do not add adversarial robustness.}, unlike in our defense where the linear layer weights are separately specified using PCA. Thus, the intution for the effectiveness of linear transformations carries over from the case of Linear SVMs, with the adversary losing access to dimensions which aid in the creation of adversarial samples, while the classifier remains largely unaffected since most of the information required for classification is retained. Further, in our empirical results in Section~\ref{sec:results}, it is clear that the average distance to the boundary of the classifier increases when the linear transformation is added, thus leading to robustness.

Finally, we note that for Linear SVMs, standard regularization can also be understood as increasing the price of all components, which reduces the dependence of the classifier on components that only provide marginal classification benefit.
However, in the case of neural networks, the usual regularization of the weight matrix does not follow the intuition given here for increasing robustness.

\subsection{Attacks against the linear transformation defense}
In order to evaluate our proposed defense mechanism, we carry out the attacks described in Section~\ref{sec:known-attacks} against classifiers learned using our defense.
Both for linear classifiers and neural networks, the classifier learned using our defense lies in the same family as the classifier that would be learned without the defense. Thus, simple modifications of existing attacks give the white-box versions of attacks against the classifier with the defense.

\subsubsection{White-box attacks} Due to the inclusion of a linear transformation of the data, the overall classifier is $f(\bfx) = f^{\text{aux}}(\bfB \bfx)$. 
In the white-box setting, since the adversary is aware of the exact parameters of the classifier produced by the defense, attacks are carried out with respect to the overall classifier. For the \emph{optimal attack on Linear SVMs}, a similar change is made, where each $\bfw_i$ (the output of the SVM optimization) is replaced by $\bfB \tran \bfw_i$, since that is the term which acts on the input $\bfx$. Thus, the adversarial sample now is
\begin{align}
\xad_t =\bfx - \frac{ (\bfz_{t,y}\tran \bfB \bfx) \bfB \tran \bfz_{t,y}}{\|\bfB \tran \bfz_{t,y}\|^2_2}.
\end{align}

In the case of \emph{gradient based attacks on neural networks}, the gradient of the loss is 
\begin{equation}
\nabla_{\bfx} J(\bfx,y,\theta) = \nabla_{\bfx} J^{\text{aux}}(\bfB\bfx,y,\theta) = \bfB \tran \nabla_{\bfx} J^{\text{aux}}(\bfx,y,\theta),
\end{equation}
where $J$ is the loss function associated with $f$ and $J^{\text{aux}}$ is associated with $f^{\text{aux}}$.
The loss of the neural network is computed with respect to its input, which is now $\bfB \bfx$, but the adversary has to add a perturbation to the input $\bfx$, which causes the largest increase in loss (up to first order). The FG attack on the defended network is then
\begin{equation}
\xad(\epsilon)= \bfx+ \epsilon   \frac{\bfB \tran \nabla_{\bfx} J^{\text{aux}}(\bfx,y,\theta)}{\| \bfB \tran \nabla_{\bfx} J^{\text{aux}}(\bfx,y,\theta) \|_2}.
\end{equation}
In the case of the \textit{optimization based attack} on neural networks, the optimization objective remains the same, with a change in the classifier the loss function is computed over:
\begin{align}
\min \quad & d(\xad, \bfx) + \lambda \ell(\xad, f),\\
\text{s.t.} \quad & \xad \in C \nonumber,
\end{align}
where $f(\bfx) = f^{\text{aux}}(\bfB \bfx)$. In our experiments, we first compute the linear transformation matrix, and then add it as a linear layer after the input layer of the neural network.

\subsubsection{Classifier mismatch attacks}
In this setting, the adversary trains a classifier $\hat{f}$ that mimics the original classifier $f$, but is not aware of the defense. 
We assume the adversary is able to train $\hat{f}$ such that it perfectly matches $f$ trained on the original data without any linear transformations.
The adversarial samples are thus generated with respect to $\hat{f} = \textsf{Train}(\bfX)$, and not with respect to the true classifier $f = (\bfx \mapsto (\textsf{Train}(\bfB\bfX)(\bfx))$.
Equivalently, this setting corresponds to the adversary using Algorithm~\ref{alg:LTtrain} with the true training examples $\bfX$, the true training function $\textsf{Train}$, but with a different choice of $\textsf{Select}$.
The adversary does not know the true $\textsf{Select}$ function, so they use a trivial version that always returns $\V{I}$.

\subsubsection{Architecture mismatch attacks}
In this setting, the adversary trains a classifier $\hat{f}$ using a choice of $\textsf{Train}$ that does not match that used to produce $f$.
Thus $\hat{f}$ is not only a different function from $f$, but it may come from a different family of classifiers.
For example, $f$ may be a three layer neural network and $\hat{f}$ may be a five layer neural network.
As in the classifier mismatch setting, the adversary is not aware of the defense used (the choice of $\textsf{Select}$).

The \emph{classifier and architecture mismatch attack settings are interesting to consider since the problem of the transferability of adversarial samples~\cite{papernot2016transferability} is still an open research question}. Our results in these settings demonstrate that a defense using linear transformations can mitigate the threat posed by transferability.

\subsubsection{Goal of the defenses}
The goal of our defense is to increase classifier robustness.
Specifically, \emph{the defenses increase the distance between benign examples and nearest adversarial examples}. Unlike some proposed defenses, we are not attempting to make the process of finding adversarial examples computationally difficult. While it is possible to use known attacks against classifiers produced by our defense, the resulting adversarial examples will be farther away from the benign examples that the adversarial examples for undefended classifiers. Thus, the fact that these attacks find adversarial examples is \emph{not} a limitation of our approach.

%%% Local Variables:
%%% mode: latex
%%% TeX-master: "NDSS"
%%% End:

\section{Experimental setup}\label{sec:exp_setup}
In this section we provide brief descriptions and implementation details of the datasets, machine learning algorithms, dimensionality reduction algorithms, and metrics used in our experiments.

\subsection{Datasets} \label{subsec:datasets}
In our evaluation, we use two datasets. The first is the MNIST image dataset and the second is the UCI Human Activity Recognition dataset.
We now describe each of these in detail. 

\subsubsection{MNIST} This is a dataset of images of handwritten digits~\cite{lecun1998mnist}. There are 60,000 training examples and 10,000 test examples. Each image belongs to a single class from 0 to 9.  The images have a dimension of $28 \times 28$ pixels (total of 784) and are grayscale. The digits are size-normalized and centred. This dataset is used commonly as a `sanity-check' or first-level benchmark for state-of-the-art classifiers.  We use this dataset since it has been extensively studied from the attack perspective by previous work. It is also easy to visualize the effects of our defenses on this dataset. 

\subsubsection{UCI Human Activity Recognition (HAR) using Smartphones} This is a dataset of measurements obtained from a smartphone's accelerometer and gyroscope~\cite{anguita2013public} while the participants holding it performed one of six activities. Of the 30 participants, 21 were chosen to provide the training data, and the remaining 9 the test data. There are 7352 training samples and 2947 test samples. Each sample has 561 features, which are various signals obtained from the accelerometer and gyroscope. The six classes of activities are Walking, Walking Upstairs, Walking Downstairs, Sitting, Standing and Laying. We used this dataset to demonstrate that our defenses work across multiple datasets and applications.

\subsection{Machine learning algorithms}\label{subsec:sim_method}
We have evaluated our defenses across multiple machine learning algorithms including linear Support Vector Machines (SVMs) and a variety of neural networks with different configurations. All experiments were run on a desktop running Ubuntu 14.04, with an 4-core Intel\textsuperscript{\tiny\textregistered} Core\textsuperscript{TM} i7-6700K CPU running at 4.00GHz, 24 GB of RAM and a NVIDIA\textsuperscript{\tiny\textregistered} GeForce\textsuperscript{\tiny\textregistered} GTX 960 GPU. 

\subsubsection{Linear SVMs}: Ease of training and the interpretability of separating hyperplane weights has led to the use of Linear SVMs in a wide range of applications~\cite{Scholkopf:2001:LKS:559923,arp2014drebin}. We use the \textsf{LinearSVC} implementation from the Python package scikit-learn~\cite{scikit-learn} for our experiments, which uses the `one-versus-rest' method for multi-class classification by default. 

In our experiments, we obtained a classification accuracy of 91.5\% for the MNIST dataset and 96.7\% for the HAR dataset. 

\subsubsection{Neural networks} Neural networks can be configured by changing the number of layers, the activation functions of the neurons, the number of neurons in each layer etc. We performed most of our experiments on a standard neural network used in previous work, for the purposes of comparison. The network we use is a standard one from~\cite{szegedy2014intriguing} which we refer to as \textsf{FC100-100-10} and a variant of it, \textsf{FC200-200-200-10}. The first neural network has an input layer, followed by 2 hidden layers, each containing 100 neurons, and an output softmax layer containing 10 neurons. Similarly, the second neural network has 3 hidden layers, each containing 200 neurons. Each neuron has a sigmoid activation function and the loss function used for training is the cross-entropy loss. We also ran experiments with a neural network with Rectified Linear Units (ReLU) as the neurons. We omit those results here due to lack of space and since they are very similar to the sigmoid activation results for the datasets we use. Both \textsf{FC100-100-10} and \textsf{FC200-200-200-10} are trained with a learning rate of 0.01 and momentum of 0.9 for 500 epochs. The size of each minibatch is 500. On the MNIST test data, we get a classification accuracy of 97.71\% for \textsf{FC100-100-10} and 98.02\% for \textsf{FC200-200-200-10}. We use Theano~\cite{theano2016}, a Python library optimized for mathematical operations with multi-dimensional arrays and Lasagne~\cite{sander_dieleman_2015_27878}, a deep learning library that uses Theano, for neural network experiments. 

Our classification accuracy results for both Linear SVMs and fully connected neural networks are comparable to baseline numbers\footnote{http://yann.lecun.com/exdb/mnist/} for corresponding architectures on MNIST, validating our implementation.

\subsection{Linear transformation techniques}

We use the PCA module from scikit-learn~\cite{scikit-learn}. 
Depending on the application, either the number of components to be projected onto, or the percentage of variance to be retained can be specified. After performing PCA on the vectorized MNIST training data to retain 99\% of the variance, the reduced dimension is 331, which is the first reduced dimension we use in our experiments on PCA based defenses. For \textit{anti-whitening}, we use a slight modification of the PCA interface to create the required transformation matrix.

\subsection{Metrics}\label{subsec:metrics}
We evaluate the relationship between $\epsilon = \|\bfx - \xad\|$, which is the allowed distance between the original example and the adversarial example, and the adversarial success rate, which we compute as follows.
For each benign input $\bfx$ with true label $y$, we check two conditions: if after perturbation, $f(\xad) \neq f(\bfx)$ \textit{and} if initially, $f(\bfx) = y$. 
Thus, the adversary's attempt is successful if the original classification was correct but the new classification on the adversarial sample is incorrect. In a sense, this count represents the number of samples that are truly adversarial, since it is only the adversarial perturbation that is causing misclassification, and not an inherent difficulty for the classifier in classifying this sample. While reporting adversarial success rates, we divide this count by the total number of benign samples correctly classified after they pass through the entire robust classification pipeline.

%%% Local Variables:
%%% mode: latex
%%% TeX-master: "NDSS"
%%% End:

\section{Experimental results}\label{sec:results}
In this section we present an overview of our empirical results. The main questions we seek to answer with our evaluations are:
\begin{enumerate}[(i)]
\item Is the defense effective in the classifier mismatch setting?
\item Is the defense effective in the white box setting?
\item Does the defense work for different classifier families?
\item Does the defense generalize across different datasets?
\item Which linear transformations are most effective?
\end{enumerate}

Our evaluation results confirm the effectiveness of our defense in a variety of scenarios, each of which has a different combination of datasets, machine learning algorithms, attacks and linear transformation used for the defense. For each set of evaluations, we vary a particular step of the classification pipeline and fix the others. Our results are summarized in Table~\ref{tbl:resultstable}.

\textbf{Baseline configuration}: We start by considering a classification pipeline with the \textit{MNIST dataset} as input data, a \textit{Linear SVM} as our classification algorithm and \textit{PCA} as the linear transformation used in our defense. Since we consider the Linear SVM as our classifier, we evaluate its susceptibility to adversarial samples generated using the \textit{attack on Linear SVMs} described in Section~\ref{subsec:attacks}. We evaluate our defenses on adversarial samples created starting from the \textit{test set} for each dataset. Unless otherwise noted, all security and utility results are for the \textit{complete} test set. To empirically demonstrate that our defense is resilient not only in this baseline case, but also various configurations of it, we systematically investigate its effect as each component of the pipeline, as well as the attacks, are changed. 

Note that in all of the plots showing the effectiveness of our defense, the legend key `None' denotes adversarial success for a classifier without any defense.

\begin{table*}{}
\centering
\begin{tabular}{c|c|c|c|c|c}
\toprule
Data set & Classifier & Attack type & Defense type and parameter & Robustness improvement & Accuracy reduction\\ \midrule
MNIST & Linear SVM & Classifier mismatch & PCA (80) & $25 \times$ & 0.22\% \\
MNIST & Linear SVM & White-box (optimal) & PCA (80) & $6 \times$ & 0.22\% \\
MNIST & \textsf{FC100-100-10} & White-box (FG) & PCA (40) & $1.5 \times$ & 0.76\% \\
MNIST & \textsf{FC100-100-10} & White-box (FGS) & PCA (40) & $2.2 \times$ & 0.76\% \\
MNIST & \textsf{FC100-100-10} & White-box (Opt.) & PCA (40) & $1.7 \times$ & 0.76\% \\
MNIST & \textsf{FC200-200-200-10} & Arch. mismatch (\textsf{FC100-100-10}) & PCA (40) & $2.4 \times$ & 0.85\% \\
MNIST & \textsf{FC100-100-10} & White-box (FG) & Anti-whiten (2) & $1.7 \times$ & 0.15\% \\
HAR & Linear SVM & White-box (optimal) & PCA (70) & $1.5 \times$ & 2.3\% \\
\bottomrule
\end{tabular}
\caption{\textbf{Robustness improvement at a misclassification rate of 60\%}. The table also gives the accuracy reduction for different classifiers, attacks and defenses on the MNIST and HAR datasets.}
\vspace{-15pt}
\label{tbl:resultstable}
\end{table*}

\begin{figure}
\centering
\input{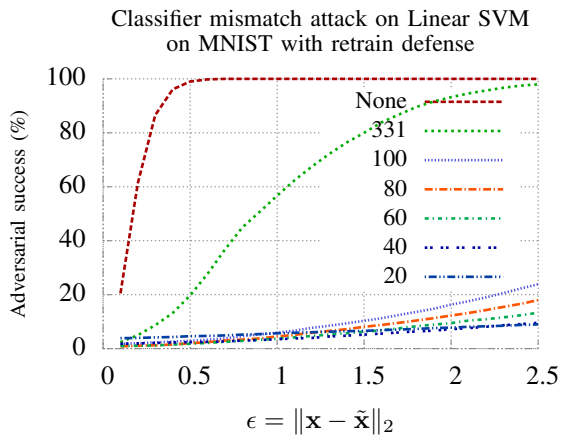}

\caption{\textbf{Effectiveness of the defense in classifier mismatch setting for the MNIST dataset with Linear SVMs}. The adversarial example success on the MNIST dataset is plotted versus the perturbation magnitude $\epsilon= \|\bfx - \xad \|_2$. The attack is performed against the original classifier and the effect of the defense is plotted for each reduced dimension $k$.}
\label{fig:svm_4}
\vspace{-15pt}
\end{figure}

\subsection{Effect of defense on Support Vector Machines} 

In the baseline case, we begin by answering questions (i) and (ii) for Linear SVMs.

\begin{figure}
  \centering
  \input{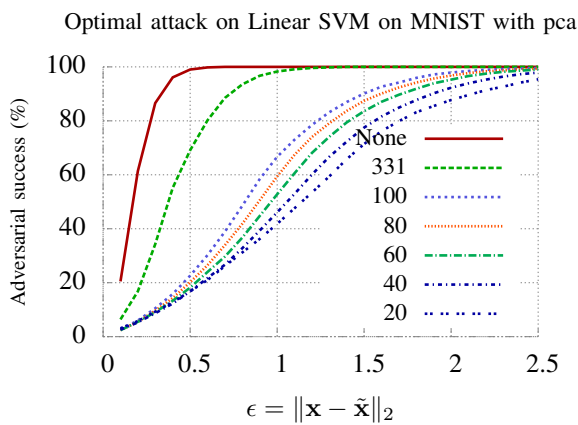}
  
  \caption{\textbf{Effectiveness of the defense for the MNIST dataset against optimal white-box attacks on Linear SVMs}. }
  %The adversarial example success on the MNIST dataset is plotted versus the perturbation magnitude $\epsilon= \|\bfx - \xad \|_2$. The attack is performed against each reduced dimension classifier and the effect of the defense is plotted.
  
  \label{fig:svm_opt}
  \vspace{-15pt}
\end{figure}

\begin{figure}
  \centering
  \input{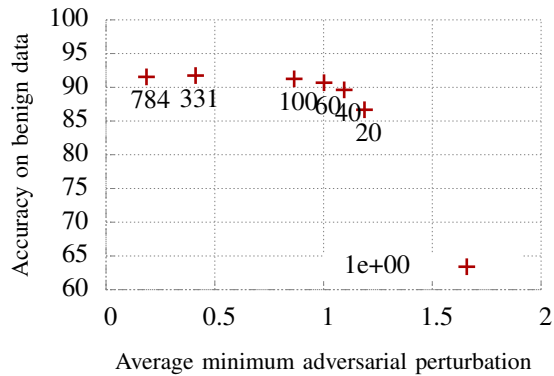}
   
  \caption{\textbf{Tradeoff between SVM classification performance on benign test data, and adversarial performance}. The x-axis represents the average over test samples of the minimum perturbation needed to cause misclassification. The legend 1e+00 denotes the regularization constant used to train the SVM.}
  \label{fig:svm-frontier}
  
\end{figure}

\subsubsection{Defense in the classifier mismatch setting}
Figure~\ref{fig:svm_4} shows the variation in adversarial success against SVMs in the classifier mismatch setting.
The defense significantly reduces adversarial success rates. For example, at $\epsilon=1.0$, the defense using PCA with a reduced dimension of $k=80$ reduces adversarial success from 100\% to 3.4\% \footnote{We note here that all changes in adversarial success (and utility) are expressed in terms of the change in \textit{percentage points}, i.e. a fall of $x \%$ indicates the absolute difference in the percentages, and not a relative difference.}. This is a 96.6\% or around $29.4\times$ decrease in the adversarial success rate. At $\epsilon=0.5$, where the adversarial success rate is 99\% , the defense with $k=80$ brings the adversarial success rate down to just $2\%$, which is a $49.5\times$ decrease. Clearly, training with reduced dimension data leads to more robust Linear SVMs, and this can be seen in the effectiveness of the defense.

Further, as we decrease the reduced dimension $k$ used in the projection step of the defense, adversarial success decreases, allowing the system designer to tune the defense according to her needs. At $k=331$, the adversarial success rate is 56.7\% at $\epsilon=1.0$, which drops to 5.9\% when $k=100$. At $k=30$, at the same $\epsilon$, the adversarial success rate drops to 4.4\% with a further decrease to 1.42\% using aggressive dimensionality reduction with $k=10$. 

In the classifier mismatch setting, the defense also acts like a noise removal process, removing adversarial perturbations and leaving behind the clean input data.
This accounts for the added robustness we see in this setting as compared to the white box setting.
Further, this \textit{mitigates the problem of the transferability of adversarial examples} when the attacker is only aware of the classifier used and not of the defense.

\subsubsection{Defense in the white box setting} 
Figure~\ref{fig:svm_opt} shows the variation in adversarial success for the defense against the \textit{optimal attack} on Linear SVMs.
This plot corresponds to the case where the adversary is aware of the dimensionality reduction defense and inputs a sample to the pipeline which is designed to optimally evade the reduced dimension classifier.
At a perturbation magnitude of 0.5, where the classifier with no defenses has a misclassification rate of 99.04\%, the reduced dimension classifier with $k=80$ has a misclassification rate of just 19.75\%, which represents a 80.25\% or $5.01 \times$ decrease in the adversarial success rates. 
At an adversarial budget of 1.3, the misclassification rate for the classifier with no defenses is 100\%, while it is about 66.7\% for the classifier with a reduced dimension of 40.

We can also study the \textit{effect of our defense on the adversarial budget required to achieve a certain adversarial success rate}. A budget of 0.3 is required to achieve a 86.6\% misclassification without the defense, while the required budget for a classifier with a defense with $k=40$ is 1.75, which is a $5.83 \times$ increase. The corresponding numbers to achieve a 98\% misclassification rate are 0.5 without the defense and 2.5 with, which represents a $5 \times$ increase. The presence of the defense forces the adversary to add much larger perturbations to achieve the same misclassification rates. Thus, our defense clearly reduces the effectiveness of an attack carried out by a very powerful \textit{adversary with full knowledge of the defense and the classifier} as well as the ability to carry out optimal attacks.

\subsubsection{Utility-security tradeoff for defense} 
Figure~\ref{fig:svm-frontier} shows the tradeoff between performance under ordinary and adversarial conditions. 
%\mn{Explain graph more clearly}
%\mn{Some points in the graph are hard to read.}
The the kink in the tradeoff for this dataset is clearly between 80 and 60. 
There is very little benefit in classification performance by using more dimensions, and essentially no benefit in robustness by using fewer. At $k=80$, we see a drop in classification success on the test set from 91.5\% without any defenses, to 90.64\% with the defense. Thus, there is a modest utility loss of about 1.2\% at this value of $k$, as compared to a security gain of $5.9 \times$, since the perturbation needed to cause 50 \% of the test set to be misclassified increases from 0.16 to 0.95.
%\mn{Caption says adversarial robustness, different from x-axis}

With these results, we can conclude that our defense is effective for Linear SVMs in both the classifier-mismatch and white-box settings. Now, we investigate the performance of our defenses on neural networks, to substantiate our claim of the applicability of our defenses across machine learning classifiers. 

\subsection{Effect of defense on neural networks} 

We now modify the baseline configuration by changing the classifier used to \textsf{FC100-100-10}. We evaluate our defenses on both gradient-based attacks for \textsf{FC100-100-10}: the Fast Gradient (FG) and Fast Gradient Sign (FGS) attacks as well as on the \textit{state-of-the-art optimization based attack} from Carlini et al.~\cite{carlini2016towards}. We continue to use the MNIST dataset and PCA as the linear transformation. With these experiments, we answer question iii), i.e. `Do the defenses work for different classifiers?' and further strengthen our claim that our defense is effective in the white box setting.

\begin{figure}
  \centering
  \input{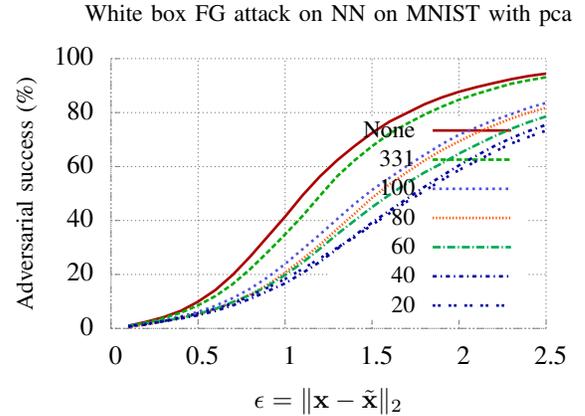}
   
  \caption{\textbf{Effectiveness of the defense for the MNIST dataset against FG attacks in the white box setting on \textsf{FC100-100-10}}.}
  
  \label{fig:nn_fg_results}
  \vspace{-15pt}
\end{figure}

\begin{figure}
  \centering
  \input{./datafigures/plots/fgs_mnist_nn_2_100_strat_pca_.tex}
   
  \caption{\textbf{Effectiveness of the defense for the MNIST dataset against FGS attacks in the white box setting on \textsf{FC100-100-10}}.}
  
  \label{fig:nn_fsg_results}
  \vspace{-15pt}
\end{figure}

\begin{figure}
  \centering
  \input{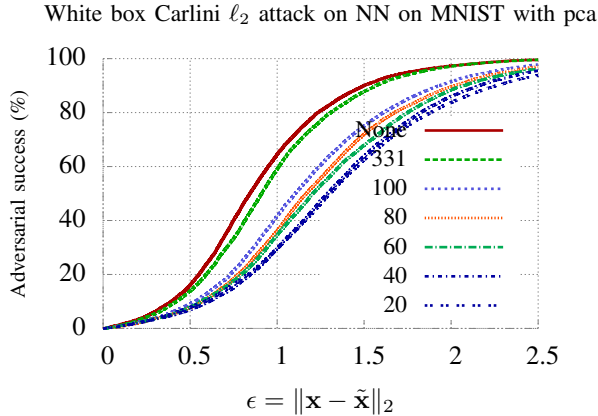}
   
  \caption{\textbf{Effectiveness of the defense for the MNIST dataset against Carlini's $\ell_2$ untargeted attack in the white-box setting on \textsf{FC100-100-10}}.}
  
  \label{fig:nn_carlini_results}
  \vspace{-15pt}
\end{figure}

\subsubsection{Defense against Fast Gradient attack in the white box setting} 
Figure~\ref{fig:nn_fg_results} shows the variation in adversarial success for the defense as $\epsilon=\| \bfx - \xad \|_2$, the parameter governing the strength of the FG attack, increases. The defense also reduces adversarial success rates for this attack. At $\epsilon=1.0$, the defense using PCA with a reduced dimension of $k=40$ reduces adversarial success from 41.42\% to 16.7\%. This is a 24.72\% or around $2.5\times$ decrease in the adversarial success rate. Again, at $\epsilon=1.5$, while the adversarial success rate is 72.42\% without any defense, the defense with $k=40$ brings the adversarial success rate down to $39.19\%$, which is a $33.23$\% or $1.8 \times$ decrease. Thus, even for neural networks, the defense causes significant reductions in adversarial success. 

Again, we can study the effect of our defense on the adversarial budget required to achieve a certain misclassification percentage. A budget of roughly $\epsilon=1.3$ is required to achieve a 60\% misclassification without the defense, while the required budget for a classifier with the defense using $k=40$ is 2, which is a $1.53 \times$ increase.

Directly comparing neural networks and linear SVMs, it appears that neural network are more robust to $\ell_2$ constrained attacks. However, it should be noted that while the Linear SVM robustness was evaluated on optimal attacks, the non-convex nature of the classification function in neural networks implies that the FG attack is only a first order approximation to an optimal attack. 

\subsubsection{Defense against Fast Gradient Sign attack in the white box setting}
The FGS attack is constrained in terms of the $\ell_\infty$ norm, so all features with non-zero gradient are perturbed by either $\eta$ or $- \eta$. 
The MNIST dataset has pixel values normalized to lie in $[0,1]$. Thus if $\eta = 0.5$, the image with every pixel equal to $0.5$ can be generated from any initial image. We restrict $\eta$ to be less than 0.25.

In Figure~\ref{fig:nn_fsg_results}, at $\eta=0.05$, the adversarial success rate falls from 41.64\% to 10.14\% for the defense with $k=40$ which is a 31.5\% or $4.1 \times$ reduction. Also, at $\eta = 0.11$, the adversarial success rate is 91.59\% without the defense and 48.92\% for the defense with $k=40$, which is a 42.67\% or $1.87 \times$ reduction. Further, the perturbation needed to cause 90\% misclassification is 0.11 without the defense but increases to 0.23 for the defense with $k=40$, which is a $2.1 \times$ increase.

\subsubsection{Defense against optimization based attack in the white-box setting}
We use Carlini and Wagner's~\cite{carlini2016towards} $\ell_2$ constrained attack to find untargeted adversarial samples, i.e. the closest possible $\xad$ in terms of the $\ell_2$ norm. Since this attack returns the minimal possible perturbation for each sample, in Figure~\ref{fig:nn_carlini_results} we plot the CDF of the minimal perturbations found by the attack over the test set in order to compare using the same metric as the other results. To see that this attack is indeed more powerful than the Fast Gradient attack (which uses the same distance metric), note that at $\|\bfx-\xad\|=1.0$, the adversarial success is around 65\% compared to 41.42\% for the FG attack, and at $\|\bfx-\xad\|=1.5$ it is 90\% compared to 72.42\% for the FG attack. 

We repeated the attack on neural networks enhanced using our PCA-based defense. The attack was carried out on the composite classifier, thus representing the white box setting. In this case, we see that at $\|\bfx-\xad\|=1.0$ the adversarial success falls to 29.5\% using the defense with $k=40$, which represents a drop of $35.5\%$ or $2.2 \times$. At a larger allowed budget of 1.5, the fall is 26.4\% or $1.4 \times$ to 63.8\%. Further, the budget required to achieve a misclassification rate of 90\% increases from 1.5 to 2.16, which is a $1.44 \times$ increase.  

\subsubsection{Defense against Fast Gradient attack in the architecture mismatch setting}
\begin{figure}
  \centering
  \input{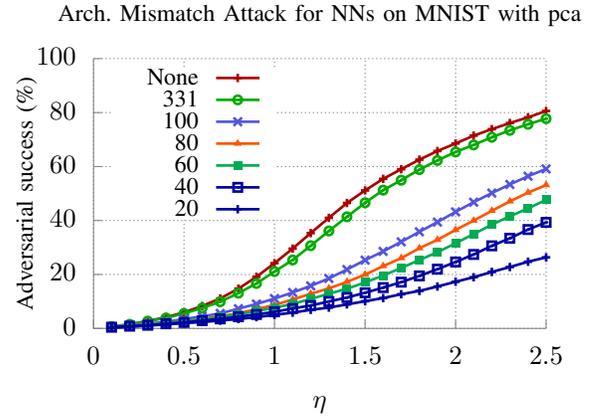}
   
  \caption{\textbf{Effectiveness of the defense for the MNIST dataset for FG samples generated for \textsf{FC100-100-10} on \textsf{FC200-200-200-10} (architecture mismatch setting)} .}
  
  \label{fig:nn_arch_mismatch_results}
  \vspace{-15pt}
\end{figure}

We now consider a setting where the adversary is less powerful. In the \emph{architecture mismatch} setting, the adversary creates adversarial sample for a different neural network (\textsf{FC100-100-10}) than the one being attacked (\textsf{FC200-200-200-10}). These results are presented in Figure~\ref{fig:nn_arch_mismatch_results}. We see a significant drop in adversarial success when our defense is used. For example, at $\|\bfx-\xad\|=1.5$ the adversarial success falls from 51.2\% to 13.2\% using the defense with $k=40$, which is a 38.0\% or $3.9 \times$ drop. Also, the budget required to achieve a misclassification rate of 40\% increases from 1.3 to 2.5, which is a $2 \times$ increase. The performance of our defense in this setting, which is commonly referred to as a `black-box' setting highlights that the defense can mitigate the transferability of adversarial samples to a large extent.

With these results for neural networks, we conclude that our defenses are effective against a variety of different attacks, in each of which the nature of the adversarial perturbation is very different. 
We have also shown that \textit{our defense is effective against the state of the art attack for neural networks in the white-box setting}, making a case for it to be included as a crucial component of any defensive measures against evasion attacks. 
These results also demonstrate that our defense can be used in conjunction with different types of classifiers, providing a general method for defending against adversarial inputs.

\subsection{Applicability for different datasets}

Next, we modify the baseline configuration by changing the datasets used. We show results with Linear SVMs as the classifier and PCA as the dimensionality reduction algorithm. We present results for the Human Activity Recognition dataset.

\subsubsection{Defense for the HAR dataset} 
\begin{figure}
  \centering
  \input{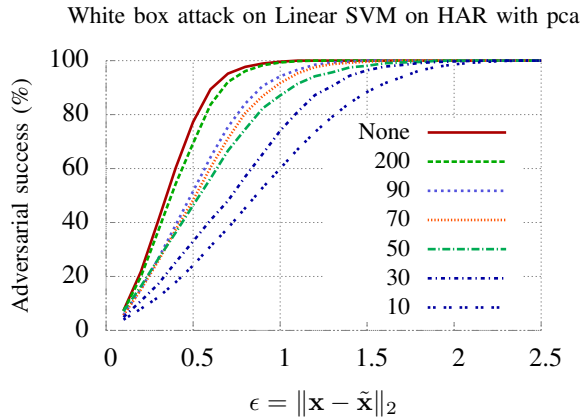}
   
  \caption{\textbf{Effectiveness of the defense for the HAR dataset against optimal white-box attacks on Linear SVMs}. Adversarial example success on the HAR dataset versus perturbation magnitude $\epsilon$ for the Linear SVM attack. Plotted for each reduced dimension $k$ used in the defense.}
  \label{fig:HAR}
\end{figure}

In Figure~\ref{fig:HAR}, the reduction in adversarial success of a white-box attack due to the defense using PCA on the HAR dataset is shown. At $\epsilon=0.5$, the adversarial success rate drops from 77.3\% without the defense to 48.3\% with $k=70$ which is a $1.6 \times$ drop respectively. In order to achieve a misclassification rate of 90\%, the amount of perturbation needed is 0.65 without the defense, which increases to 0.93 with $k=70$. Thus, the adversarial budget increases $2 \times$ to achieve the same adversarial success rate. The impact on utility is modest: a drop of 2.3\% for $k=70$, which is small in comparison to the gain in security.

\subsection{Effect of PCA-based defense on utility}
Table~\ref{tbl:utiltable} presents the effect of our defense on the classification accuracy of benign data. The key takeaways are that the decrease in accuracy for both neural networks and Linear SVMs for reduced dimensions down to $k=50$ is at most 4\%. Further, we notice that dimensionality reduction using PCA can actually \textit{increase} classification accuracy as when $k=70$, the accuracy on the MNIST dataset increases from 97.47\% to 97.52\%. This effect probably occurs as the dimensionality reduction can prevent over-fitting. More aggressive dimensionality reduction however, can lead to steep drops in classification accuracy, which is to be expected since much of the information used for classification is lost.

\begin{table}
\centering
\begin{tabular}{@{}ccccc}\toprule
\multicolumn{3}{c}{MNIST data} & \multicolumn{2}{c}{HAR data}\\
\cmidrule(lr){1-3} \cmidrule(lr){4-5} 
$k$ & \textsf{FC100-100-10} & Linear SVM & $k$  & Linear SVM  \\ 
\cmidrule(lr){1-3} \cmidrule(lr){4-5} %\midrule
No D.R. & 97.47 & 91.52 & No D.R. & 96.67 \\
784 & 97.32 & 91.54 & 561 & 96.57 \\
					
331 & 97.35 & 91.37 & 200 & 96.61 \\

100 & 97.36 & 90.89 & 90 & 94.60 \\

80 & 97.25 & 90.64 & 70 & 94.37 \\

60 & 97.38 & 90.47 & 50 & 92.47 \\

40 & 96.71 & 89.03 & 30 & 91.11 \\

%20 & 96.67 & 86.69 & 10 & 86.67 \\ 
\bottomrule
\end{tabular}
\caption{\textbf{Utility values for the dimensionality reduction defense}. For the MNIST and HAR datasets, the classification accuracy on the benign test set is provided for various values of reduced dimension $k$ used for the PCA based defense, as well as the accuracy without the defense.}

\label{tbl:utiltable}
\vspace{-15pt}
\end{table}

\subsection{Defense using anti-whitening}
\begin{figure}
  \centering
  \input{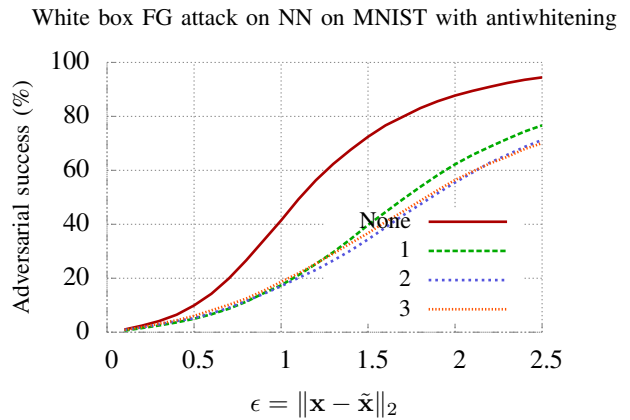}
  
  \caption{\textbf{Effectiveness of the anti-whitening defense for the MNIST dataset against FG attacks in the white box setting on \textsf{FC100-100-10}}.}
  
  \label{fig:nn_fg_antiwhiten_results}
\end{figure}

As described in Section~\ref{subsec: antiwhiten}, anti-whitening is a soft approximation of PCA where high-variance components are boosted with respect to the low-variance ones, instead of just dropping them. 
This can be controlled by the parameter $c$ in the anti-whitening transformation $\bfB = \Lambda^{\frac{c}{2}} \bfU^T$. 
In Figure~\ref{fig:nn_fg_antiwhiten_results}, the effects of the defense using anti-whitening with $c=1,2$ and 3 are shown. At $\epsilon=1.0$, the defense with $c=2$ causes the adversarial success to fall from 41.42\% to 17.06\%, which is a 24.36\% or $2.4 \times$ fall. 
At $\epsilon = 1.5$, the corresponding reduction is from 72.42\% to 34.42\%, which is a 38\% or $2.1 \times$ decrease. 
The anti-whitening defense thus performs slightly better than the PCA defense with a comparable parameter ($k=40$).\\
\textit{Effect of anti-whitening on utility}: For $\textsf{FC100-100-10}$, the classification rate on benign data is 97.47\% without any defense. Using anti-whitening with $c=1,2$ and 3, the utility values are 97.45\%, 97.32\% and 96.83\% respectively. This shows that the anti-whitening defense is slightly better both with respect to both security and utility as compared to the PCA defense. The increase in utility is likely due to the fact that dimensions are not dropped and are used to achieve better classification performance. 

These results highlight the broad applicability of our defense across application domains. It is clear that the effectiveness of our defense is not an artifact of the particular structure of data from the MNIST dataset, and that the intuition for its effect holds across different kinds of data.

%%% Local Variables:
%%% mode: latex
%%% TeX-master: "NDSS"
%%% End:

\section{Discussions, Limitations and Future Work}\label{sec:discussion}

Even though our defense reduces adversarial success rates and increases the amount of perturbation the adversary has to add to achieve fixed levels of misclassification in a number of cases, there are two main areas where it can be improved in conjunction with other defense mechanisms. 
%it falls short of being a comprehensive defense mechanism against evasion attacks:

\subsubsection{Further reductions in adversarial success} While our defense causes 
significant reductions in adversarial success rates in a variety 
of settings, there are cases where the adversarial success rate 
is still non-trivial. In such cases, it is likely our defense would 
have to be combined with other defenses such as adversarial 
training~\cite{goodfellow2014explaining} and ensemble 
methods~\cite{SmutzS16} for detection in order to create a ML system 
secure against evasion attacks. Our defense has the advantage that 
it can be used in conjunction with a variety of ML classifiers and it 
will not interfere with the operation of other defense mechanisms. Since our defense increases the amount of perturbation needed to achieve a fixed misclassification rate, it may aid detection based defenses.

\subsubsection{Better data transformations} In certain settings, using PCA 
for dimensionality reduction may have limited applicability. For example, 
we found that our PCA based defense offers only marginal security improvement 
for the \textsf{Papernot-CNN} (See Section~\ref{subsec:app_cnns} for details).
%\mn{Instead of CNN figure, can highlight a more positive result} 
%It is likely that this effect stems from the fact that CNNs already 
%incorporate domain-specific knowledge in their convolutional layers, and an 
%additional layer of pre-processing using PCA does not confer any additional 
%robustness.
It is likely that this effect stems from PCA reducing the amount of 
local information that the convolutional layers in the CNN are able 
to use for the purposes of classification. A key step in addressing this
limitation of our defense is to use other dimensionality reduction techniques 
which could reduce adversarial success to negligible levels and work better when combined with classifiers such as CNNs. This limitation also prevents us from achieving state-of-the-art accuracy on image datasets like MNIST, since the best classifiers in for these datasets use convolutional layers. In future work we plan to explore techniques such as autoencoders and kernel PCA for designing robust classifiers. For certain problems, it may also be feasible to explicitly optimize for the linear transformation achieving the best utility-security tradeoff. This is another direction we plan to explore.
% and various compression schemes to better understand 
%the relationship between dimensionality reduction and the robustness of classifiers.

%%% Local Variables:
%%% mode: latex
%%% TeX-master: "NDSS"
%%% End:

\section{Related work} \label{sec:related}
Previous defenses against adversarial examples have largely focused on specific classifier families or application domains. Further, the existing defenses provide improved security only against existing attacks in the literature, and it is unclear if the defense mechanisms will be effective against adversaries with knowledge of their existence, i.e. strategic attacks exploiting weaknesses in the defenses. As a case in point, Papernot et al.~\cite{papernot2015distillation} demonstrated a defense using distillation of neural networks against the Jacobian-based saliency map attack~\cite{papernot2016limitations}. However, Carlini et al.~\cite{carlini2016towards} showed that a modified attack negated the effects of distillation and made the neural network vulnerable again. Now, we give an overview of the existing defenses.

\subsubsection{Classifier-specific} Russu et al.~\cite{russu2016secure} propose defenses for SVMs by adding various kinds of regularization. Kantchelian et al.~\cite{kantchelian2016evasion} propose defenses against optimal attacks designed specifically for tree-based classifiers. Existing defenses for neural networks~\cite{gu2014towards,shaham2015understanding,zhao2016suppressing,luo2015foveation,huang2015learning} make a variety of structural modifications to improve resilience to adversarial examples. These defenses do not readily generalize across classifiers and may still be vulnerable to adversarial examples, as shown by Gu and Rigazio~\cite{gu2014towards}.

\subsubsection{Application-specific} Hendrycks and Gimpel ~\cite{hendrycks2016visible} study transforming images from the RGB space to YUV space to enable better detection by humans and decrease misclassification rates. They also use whitening to make adversarial perturbations in RGB images more visible to the human eye. The effect of JPG compression on adversarial images has also been studied~\cite{dziugaite2016study,das2017keeping}. Their conclusions were that it has a small beneficial effect when the perturbations are small. These approaches are restricted to combating evasion attacks on image data and do not generalize across applications. Further, it is unclear if they are effective against white-box attacks.

\subsubsection{General defenses} An ensemble of classifiers was used by Smutz and Stavrou~\cite{SmutzS16} to \textit{detect} evasion attacks, by checking for disagreement between various classifiers. However, an ensemble of classifiers may still be vulnerable to adversarial examples since they generalize across classfiers. Further, Goodfellow et. al.~\cite{goodfellow2014explaining} show that ensemble methods have limited effectiveness for evasion attacks against neural networks. Goodfellow et. al.~\cite{goodfellow2014explaining}, Tram\`{e}r et al.~\cite{tramer2017ensemble} and M\k{a}dry et al.~\cite{madry_towards_2017} re-train on adversarial samples of different types to improve the resilience of neural networks. They all find that adversarial training works, but needs high capacity classifiers to be effective, and further, its effectiveness reduces as the perturbation is increased beyond the one used for training. In our experiments, we find that re-training on adversarial samples has an extremely limited effect on increasing the robustness of linear SVMs (see Figure~\ref{fig:svm_adv} in the Appendix), thus this defense may not be applicable across classifiers, and does indeed depend on the capacity of the classifier. Wang et al.~\cite{wang2016random} use random feature nullification to reduce adversarial success rates for evasion attacks on neural networks. The applicability of this idea across classifiers is not studied. Zhang et al.~\cite{zhang2016adversarial} use adversarial feature selection to increase the robustness of SVMs. They find and retain features that decrease adversarial success rates. This defense may be generalized across other classifiers and is an interesting direction for future work.

Due to the classifier and dataset-agnostic nature of our defense, it may be combined with existing defenses such as adversarial training which have differnet aims for an even larger improvement in robustness. For example, neural networks may be trained with reduced dimension samples, and the training process can also incorporate the adversarial loss to further increase the robustness of the network. We plan to explore these directions in future work.

%%% Local Variables:
%%% mode: latex
%%% TeX-master: "NDSS"
%%% End:

\section{Conclusion}\label{sec:conclusion} 
In this paper, we considered the novel use of data transformations such as dimensionality reduction as a defense mechanism against evasion attacks on ML classifiers. Our defenses rely on the insight that (a) linear transformations of data allow access to usually inaccessible security-performance tradeoffs, and (b) training classifiers on reduced dimension data leads to enhanced resilience of ML classifiers (by reducing the weights of less informative and low-variance features). 
Using empirical evaluation on multiple real-world datasets, we demonstrated a 2x reduction in adversarial success rates across a range of attack strategies (including white-box ones), ML classifiers, and applications. 
Our defenses have a modest impact on the utility of the classifiers (0.5-2\% reduction), and are computationally efficient.
Our work thus provides an attractive foundation for countering the threat of evasion attacks. 

%%% Local Variables:
%%% mode: latex
%%% TeX-master: "NDSS"
%%% End:

\bibliographystyle{IEEEtranS}
\bibliography{bib_adv_ml.bib}

\section{Appendix}\label{sec:appendix}

\begin{figure}
\centering

  	\includegraphics[scale=2.0]{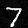}
  	\includegraphics[scale=2.0]{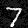}
  	\includegraphics[scale=2.0]{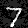}
  	\includegraphics[scale=2.0]{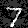}
  	\includegraphics[scale=2.0]{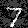}

\caption{\textbf{Adversarial images of digit `7' (against a neural network with no defense)}: The images have been modified with the Fast Gradient Sign attack on neural networks with (from left to right), $\eta \approx 0.05,0.1,0.15,0.2$ and $0.25$. The perturbation begins to be visible at $\eta=0.15$ and is very obvious in the images with $\eta>0.2$. The attack was carried out on a classifier $f$ without any dimensionality reduction.}
\label{fig:adv_fsg_images}
\end{figure}

\subsection{Complexity Analysis of PCA Defenses}\label{sec:complexity}

The defense using PCA adds a one-time $\mathcal{O}(d^2n+d^3)$ overhead for finding the principal components, with the first term arising from the covariance matrix computation and the second term from the eigenvector decomposition. There is also a one-time overhead for training a new classifier on the reduced dimension data. The time needed to train the new classifier will be less than that needed for the original classifier since the dimensionality of the input data has reduced. Each subsequent input will incur a $\mathcal{O}(dk)$ overhead due to the matrix multiplication needed to project it onto the principal components.

\begin{figure}
  \centering
  \input{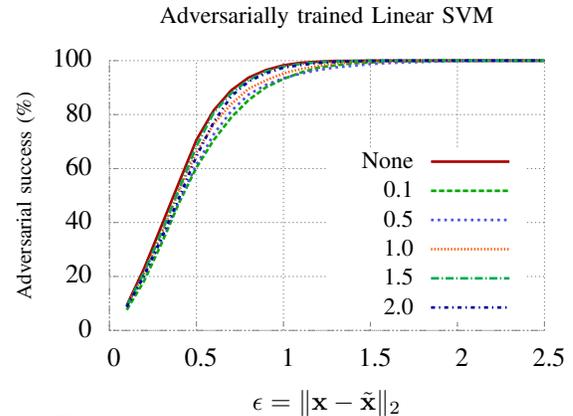}
  \vspace{-10pt}
  \caption{\textbf{Effectiveness of adversarial training for the MNIST dataset against optimal white box attacks on Linear SVMs.} The Linear SVM was trained using gradient descent with periodically augmented training sets containing adversarial samples with the specified perturbation values.}
  \vspace{-15pt}
  \label{fig:svm_adv}
\end{figure}

\subsection{CNNs}\label{subsec:app_cnns}
We also run our experiments on a Convolutional Neural Network~\cite{goodfellow2016deep} whose architecture we obtain from Papernot et al.~\cite{papernot2015distillation}. This CNN's architecture is as follows: it has 2 convolutional layers of 32 filters each, followed by a max pooling layer, then another 2 convolutional layers of 64 filters each, followed by a max pooling layer. Finally, we have two fully connected layers with 200 neurons each, followed by a softmax output with 10 neurons (for the 10 classes in MNIST). All neurons in the hidden layers are ReLUs. We call this network \textsf{Papernot-CNN}. It is trained with a learning rate of 0.1 (adjusted to 0.01 for the last 10 epochs) and momentum of 0.9 for 50 epochs. The batchsize is 500 samples for MNIST and on the MNIST test data we get a classification accuracy of 98.91\% with the \textsf{Papernot-CNN} network.

\end{document}